\documentclass[aps,prl,reprint,nofootinbib,twocolumn,superscriptaddress,showpacs,showkeys,longbibliography]{revtex4-1}
\usepackage{eurosym}
\usepackage{amsmath,amssymb,amstext}
\usepackage{ulem}
\usepackage[usenames,dvipsnames]{color}
\usepackage{graphicx}
\usepackage{braket}
\usepackage{natbib}
\usepackage{comment}
\usepackage{dcolumn}
\usepackage{wasysym}

\usepackage[colorlinks,bookmarks=false,citecolor=blue,linkcolor=red,urlcolor=blue]{hyperref}

\begin{document}
\title{Dark soliton of polariton condensates under nonresonant $\mathcal{P}\mathcal{T}$- symmetric pumping}
\author{Chunyu Jia}
\affiliation{Department of Physics, Zhejiang Normal University, Jinhua, 321004, China}
\author{Zhaoxin Liang}
\email{The corresponding author: zhxliang@gmail.com}
\affiliation{Department of Physics, Zhejiang Normal University, Jinhua, 321004, China}

\date{\today}

\begin{abstract}
A quantum system in complex potentials obeying parity-time ($\mathcal{P}\mathcal{T}$) symmetry could exhibit all real spectra, starting out in non-Hermitian quantum mechanics.
The key physics behind a $\mathcal{P}\mathcal{T}$- symmetric system consists of the balanced gain and loss of the complex potential. In this work, we plan to include the nonequilibrium nature, i.e. the intrinsic kinds of gain and loss of a system, to a $\mathcal{P}\mathcal{T}$- symmetric many-body quantum system, with the emphasis on the combined effects of non-Hermitian due to nonequilibrium nature and $\mathcal{P}\mathcal{T}$ symmetry in determining the properties of a system. In this end, we investigate the static and dynamical properties of a dark soliton of a polariton Bose-Einstein condensate under the $\mathcal{P}\mathcal{T}$- symmetric non-resonant pumping by solving the driven-dissipative Gross–Pitaevskii equation both analytically and numerically. We derive the equation of motion for the center of mass of the dark soliton’s center analytically with the help of the Hamiltonian approach. The resulting equation captures
how the combination of the open-dissipative character and $\mathcal{P}\mathcal{T}$- symmetry affects the properties of dark soliton, i.e. the soliton  relaxes
by blending with the background at a finite time.  Further numerical solutions are in excellent agreement with the analytical results.
\end{abstract}

\maketitle

At present, there are significant interests and ongoing efforts in investigating parity-time ($\mathcal{P}\mathcal{T}$)- symmetric non-Hermitian quantum mechanics~\cite{Bender1998,Bender2007,PTExp1,PTExp2,Li2019} in different contexts of physical systems. The motivation behind this research line is twofold. First, the introduction of $\mathcal{P}\mathcal{T}$- symmetry has gone conceptually beyond the Hermitian quantum mechanics. Traditionally, it was well believed that any physical phenomena in quantum systems must be described by Hermitian Hamiltonians with real eigenvalues. However, non-Hermitian systems with the $\mathcal{P}\mathcal{T}$- symmetic complex potential can have real spectra, staring out in non-Hermitian quantum mechanics~\cite{Bender2007,Makris2011}. Second, the physical systems with the $\mathcal{P}\mathcal{T}$- symmetry can provide some enlightening applications, such as single-mode $\mathcal{P}\mathcal{T}$ lasers~\cite{Hodaei2014,Peng2014} and unidirectional reflectionless $\mathcal{P}\mathcal{T}$-symmetric meta material at optical frequencies~\cite{Feng2013}. Up to now, extension of $\mathcal{P}\mathcal{T}$ symmetry to other branches of physics is a hot topic, such as mechanical oscillators~\cite{Bender2013}, optical waveguides~\cite{Regensburger2012}, and optomechanical systems~\cite{Xu2016}.

Along this research line, there are timely efforts of extending the studies of $\mathcal{P}\mathcal{T}$- symmetry to non-equilibrium quantum systems with emphasis on the combined effects of $\mathcal{P}\mathcal{T}$- symmetry and  the intrinsic non-equilibrium nature on a quantum system. In this end, Bose-Einstein condensates (BECs) of polaritons  in quantum well semiconductor microcavities~\cite{Rev2,Rev0,Rev1,Rev3} has opened up intriguing possibilities to explore $\mathcal{P}\mathcal{T}$- symmetric non-Hermitian quantum mechanics beyond thermal equilibrium. On the one hand, as polaritons undergo rapid radiative decay, their population in the condensate is maintained by persistent optical pumping. Hence a polariton condensate is inherently in non-equilibrium with an
open-dissipative character. Its mean-field physics can be well captured by a Gross–Pitaevskii equation (GPE) with gain and loss~\cite{Xue2014,Smirnov2014,Pinsker2014,Pinsker2015,Pinsker2016,Ma2017,Ma2018,Xu2019,Amo2011,Cilibrizzi2014,Walker2017}, where the nonlinearity results from the strongly and repulsively interacting excitons. On the other hand, there are ongoing experimental investigations  on non-Hermitian $\mathcal{P}\mathcal{T}$- symmetric systems in the context of polariton condensates~\cite{Gao2015,Gao2018}. For example, the non-trivial topology
of eigenmodes and unusual transport properties in the vicinity of exceptional points are currently under investigation~\cite{Gao2018}. Moreover,  the direct observation of non-Hermitian degeneracies in a chaotic exciton-polariton billiard has been done in Ref.~\cite{Gao2015}.

In this work, we propose and investigate an alternative route to study non-equilibrium $\mathcal{P}\mathcal{T}$- symmetric many-body quantum system in the context of a polariton condensate with $\mathcal{P}\mathcal{T}$- symmetric non-resonant pumping. In more details, we focus on the dynamics of a dark soliton in a polariton codensate under the nonresonant spatial dependent pumping with $\mathcal{P}\mathcal{T}$- symmetry by solving open-dissipative GP equation coupled to a rate equation for a density with a combination of analytical and numerical approaches. In this end, we first use the Hamiltonian approach to derive the equation of motion for the soliton parameters analytically, i.e. the velocity of the dark soliton. Then  we compare these analytical results with the numerical solutions for the trajectory of dark solitons by solving the GP equation. We find a remarkable agreement between the two as it's expected. Our work would open a new perspective of including the nonequilibrium nature, i.e. the intrinsic kinds of gain and loss of a system, to a $\mathcal{P}\mathcal{T}$- symmetric many-body quantum system.

\begin{figure}
  \includegraphics[width=0.85\columnwidth]{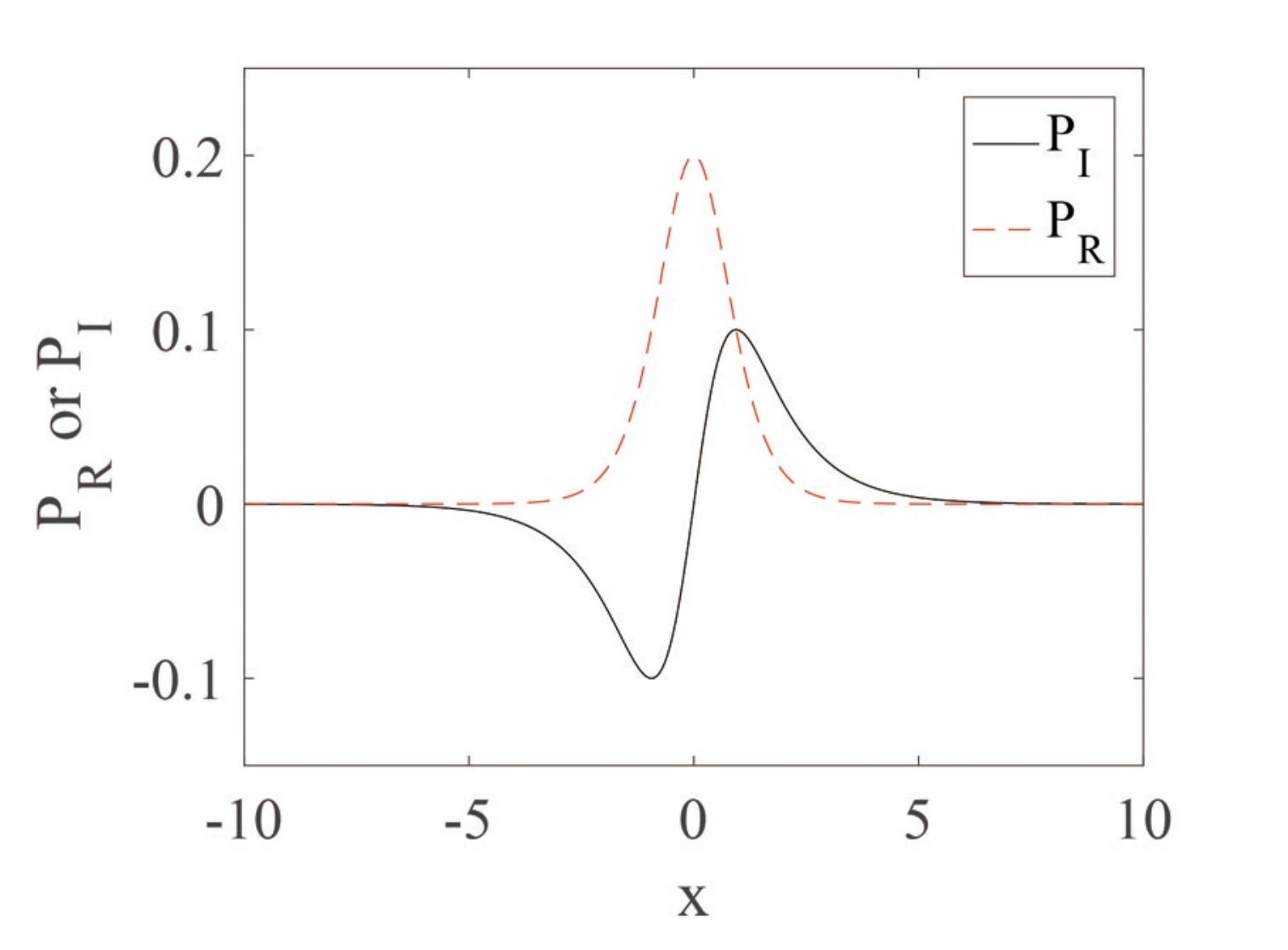}\\
\caption{The schematic profiles of the real and imaginary parts of the non-resonant PT symmetry pumping based on Eq. (\ref{PTpotential}).
}\label{Fig1}
\end{figure}

We are interested in the dynamics of the polariton condensate formed under uniform non-resonant pumping in a wire-shaped microcavity, as motivated by Ref.~\cite{Wertz2010}.
At the mean field level, the condensate can be well
described by the driven-dissipative GPE characterized by the time-dependent condensate order parameter of $\psi(x,t)$, coupled to a rate
equation for a density, $n_{R}\left(x,t\right)$, of an uncondensed
reservoir of high-energy near-excitonic polaritons as follows ~\cite{Xue2014,Smirnov2014,Pinsker2014,Pinsker2015,Pinsker2016,Ma2017,Ma2018,Xu2019}
\begin{eqnarray}
i\hbar\frac{\partial\psi\left(x,t\right)}{\partial t}&=&\Big[-\frac{\hbar^{2}}{2m_{\text{eff}}}\nabla^{2}+g_{C}\left|\psi\right|^{2}+g_{R}n_{R}\left(x,t\right)\nonumber\\
&+&\frac{i\hbar}{2}\left(Rn_{R}\left(x,t\right)-\gamma_{C}\right)\Big]\psi\left(x,t\right),\label{GPE}\\
\frac{\partial n_{R}\left(x,t\right)}{\partial t}&=&P\left(x\right)-\left(\gamma_{R}+R\left|\psi\right|^{2}\right)n_{R}\left(x,t\right).\label{Rate}
\end{eqnarray}
In Eqs. (\ref{GPE}) and (\ref{Rate}),  $\psi(x,t)$ and $n_{R}\left(x,t\right)$ represent the order parameter of the polariton condensate and
the reservoir density respectively. Here, $m_{\text{eff}}=10^{-4}m_{\text{e}}$ labels the effective mass of polaritons with $m_{\text e}$ being the free electron mass.
The $g_{C}$ and $g_{R}$ characterize the strength of nonlinear interaction of polaritons and the interaction strength between reservoir and polariton respectively. The condensed polaritons with a finite lifetime $\gamma_{C}^{-1}$ are continuously
replenished from reservoir polaritons at a rate $R$. High energy exciton-like polaritons are injected into the reservoir by laser pump
$P\left(x\right)$ and relax at the reservoir loss rate $\gamma_{R}$.

In this work, we consider that the $\mathcal{P}\mathcal{T}$- symmetric pumping of $P\left(x\right)$~\cite{Rev2,Rev0,Rev1,Rev3} in Eq. (\ref{Rate}) consist of a constant part $P_0$ and the spatial dependent part as
$P\left(x\right)=P_0+P_{{\text {$\mathcal{P}\mathcal{T}$}}}\left(x\right)$. For the pumping rate $P$ to be $\mathcal{P}\mathcal{T}$ symmetric, $P_{\text{$\mathcal{P}\mathcal{T}$}}$ must be successive action of the parity (P) and time-reversal (T) operator: $\mathcal{P}\mathcal{T}\{P_{\text{$\mathcal{P}\mathcal{T}$}}(x)\}=P_{\text{$\mathcal{P}\mathcal{T}$}}^*\left(-x\right)=P_{\text{$\mathcal{P}\mathcal{T}$}}(x)$. In more details, we are interested in $P_{\text{$\mathcal{P}\mathcal{T}$}}(x)$ in the following form
\begin{eqnarray}
P_{\text{$\mathcal{P}\mathcal{T}$}}&=& P_{\text{R}}+iP_{\text{I}}\nonumber\\
&=&V_0\text{sech}^2\left(\frac{U\xi}{x_h}\right)+iV_1\text{sech}\left(\frac{U\xi}{x_h}\right)\tanh\left(\frac{U\xi}{x_h}\right).\label{PTpotential}
\end{eqnarray}
with $V_0$ and $V_1$ being the amplitudes of the real and imaginary part of nonresonant $\mathcal{P}\mathcal{T}$- symmetric pumping and $\xi=x-v_st$, $x_h$ being the healing length, and $U=\sqrt{1-v_s^2}$. As shown in Fig. \ref{Fig1}, we plot the real part and imaginary part of $P_{\text{$\mathcal{P}\mathcal{T}$}}$, corresponding the even and odd functions respectively. Note that the imaginary part represents the gain-loss effects due to the $\mathcal{P}\mathcal{T}$- symmetric pumping.
We remark that the reservoir density $n_R$ of Eq. (\ref{Rate}) is supposed to be complex under the $\mathcal{P}\mathcal{T}$- symmetric pump, which will challenge the physical explanation of $n_R$. Our strategy is to obtain an effective equation of $\psi$ (see Eq. (\ref{FinalPT}) below) by replacing the $n_R$ in Eq. (\ref{GPE}) in the fast reservoir limit.

The emphasis and value of this work is to include the nonequilibrium nature characterized the intrinsic kinds of gain and loss of a system, to a $\mathcal{P}\mathcal{T}$- symmetric many-body quantum system, with the emphasis on the combined effects of non-Hermitian due to both nonequilibrium nature and $\mathcal{P}\mathcal{T}$ symmetry in determining the properties of a system. It's clear now that the intrinsic kinds of gain and loss of a polariton condensate are described by the parameters of $R$ and $\gamma_C$ in Eq. (\ref{GPE}); in contrast, the imaginary part of $\mathcal{P}\mathcal{T}$- symmetric pumping in Eq. (\ref{PTpotential}) captures the key physics of $\mathcal{P}\mathcal{T}$ symmetry inducing gain-loss effects of the polariton condensate. Hence, the competition of non-Hermitian due to both nonequilibrium nature and $\mathcal{P}\mathcal{T}$ symmetry in determining the properties of a system can be determined by the rich interplay among four parameters of $R$, $\gamma_C$ in Eq. (\ref{GPE}) and $V_0$, $V_1$ in Eq. (\ref{PTpotential}). In what follows, we focus on the static and dynamical properties of a dark soliton of a polariton condensate under the $\mathcal{P}\mathcal{T}$- symmetric non-resonant pumping by solving a mean-field driven-dissipative GP description both analytically and numerically.

For convenience of later analysis, we proceed to recast Eqs. (\ref{GPE}) and (\ref{Rate}) into a dimensionless
form by rescaling space time in the units of healing length $r_{h}=\hbar/\left(mc_{s}\right)$
and $\tau_{0}=r_{h}/c_{s}$. Here $c_{s}=\left(g_{C}n_{C}^{*}/m\right)^{1/2}$ is
a local sound velocity in the condensate and $n_{C}^{*}$ is a characteristic
value of the condensate density. For a cw background, it is convenient to choose $n_{C}^{*}=n_C^0=(P_0-P_{\text{th}})/\gamma_C$ with $P_{\text{th}}=\gamma_R\gamma_C/R$. The dimensionless equations for the
normalized condensate wave function $\bar{\psi}=\psi/\sqrt{n_{C}^{*}}$
,and reservoir density $\bar{n}_{R}=n_{R}/n_{C}^{*}$ take the form
\begin{eqnarray}
i\frac{\partial\bar{\psi}}{\partial t}&=&\left[-\frac{1}{2}\frac{\partial^{2}}{\partial x^{2}}+\left|\bar{\psi}\right|^{2}+\bar{g}_{R}\bar{n}_{R}+\frac{i}{2}\left(\bar{R}\bar{n}_{R}-\bar{\gamma}_{C}\right)\right]\bar{\psi},\label{Dpsi}\\
\frac{\partial\bar{n}_{R}}{\partial t}&=&\bar{P}_0+\bar{P}_{\text{$\mathcal{P}\mathcal{T}$}}(x)-\left(\bar{\gamma}_{R}+\bar{R}\left|\bar{\psi}\right|^{2}\right)\bar{n}_{R},\label{DnR}
\end{eqnarray}
In above, the dimensionless parameters are defined as follows: $\bar{g}_{R}=\frac{g_{R}}{g_{C}}$,
$\bar{R}=\frac{\hbar R}{g_{C}}$, $\bar{\gamma}_{C}=\frac{\hbar\gamma_{C}}{g_{C}n_{C}^{*}}$, $\bar{\gamma}_{R}=\frac{\hbar\gamma_{R}}{g_{C}n_{C}^{*}}$, and
$\bar{P}\left(r\right)=\bar{P}_0+\bar{P}_{\text{$\mathcal{P}\mathcal{T}$}}(x)=\frac{\hbar P\left(x,t\right)}{g_{C}n_{C}^{*2}}$.

The goal of this work is to investigate the dynamics of the dark soliton of the polariton condensate under incoherent $\mathcal{P}\mathcal{T}$-symmetric pumping.
Generally speaking, a dark soliton is referred to as the finite amplitude collective excitation of a
homogeneous condensate. Inspired by this physical picture of a dark soliton, we consider perturbations of the
condensate wave function and the reservoir density in the
following general form as Refs. \cite{Smirnov2014,Xu2019}
\begin{eqnarray}
\bar{\psi}(x,t)&=&\psi\left(x,t\right)\exp[-i(1+\bar{g}_R\bar{\gamma}_C/\bar{R})t],\label{Ppsi} \\
\bar{n}_{R}(x,t)&=&\bar{g}_R\bar{\gamma}_C/\bar{R}+m_{R}\left(x,t\right).\label{PmR}
\end{eqnarray}
By plugging Eqs. (\ref{Ppsi}) and (\ref{PmR}) into Eqs. (\ref{Dpsi}) and (\ref{DnR}), we can obtain that the perturbations $\psi\left(x,t\right)$ and $m_{R}\left(x,t\right)$
are governed by the dynamical equations
\begin{eqnarray}
i\frac{\partial \psi}{\partial t}&=&\left[-\frac{1}{2}\nabla^{2}+\left|\psi\right|^{2}-1+\bar{g}_{R}m_{R}+\frac{i}{2}\bar{R}m_{R}\right]\psi,\label{Fpsi}\\
\frac{\partial m_{R}}{\partial t}&=&\bar{P}-\left(\bar{\gamma}_{R}+\bar{R}\left|\psi\right|^{2}\right)m_{R}+\bar{\gamma}_{C}\left(1-\left|\psi\right|^{2}\right).\label{FmR}
\end{eqnarray}
Blow, we plan to present a detailed analysis on the dynamics of the dark soliton by solving Eqs. (\ref{Fpsi}) and (\ref{FmR}) analytically and numerically.

Before investigating the effects of $\mathcal{P}\mathcal{T}$- symmetric pumping on the soliton, we first briefly review some important features of a dark soliton in
the nonlinear Schr\"odinger equation, corresponding to $m_{R}\left(x,t\right)=0$ in Eq. (\ref{Fpsi}), i.e. in the absence of the open-dissipative and $\mathcal{P}\mathcal{T}$- symmetric pumping effects.
In such, there exists an exact dark-soliton solution, which takes the form
\begin{equation}
\psi_{s}(\xi=x-\upsilon_{s}t)=\sqrt{1-\upsilon_{s}^{2}}\tanh(\sqrt{1-\upsilon_{s}^{2}}\xi)+i\upsilon_{s},\label{ansatz}
\end{equation}
with $\upsilon_{s}$ being the velocity of the traveling dark soliton. In particular, for a moving soliton, the minimum value of density $n_{C}^{min}=\left|\psi_{s}\left(\xi=0\right)\right|^{2}$
increases in proportion to the square of the soliton velocity $n_{C}^{min}=\upsilon_{s}^{2}$.

Next, we proceed to include both the open-dissipative effects and $\mathcal{P}\mathcal{T}$- symmetric pumping as captured by $m_{R}\left(x,t\right)\neq 0$. In such a case, Equation (\ref{ansatz}) is not the exact solution of Eq. (\ref{Fpsi}) any more. Directly following Refs. \cite{Smirnov2014,Xu2019,Qi2019}, we plan to adopt the Hamiltonian approach for the perturbation theory of soliton. At the heart of the Hamiltonian approach of quantum dynamics for a dark soliton is based on that, in presence of perturbation, the dark soliton's velocity become slow functions of time, but the functional form of the dark soliton remains unchanged, i.e. $v_s\rightarrow v_s(t)$ in Eq. (\ref{ansatz}). Then, the time evolution of the parameter of $v_{\text{s}}$ can be routinely determined by calculating
the time variation of the dark soliton's energy
\begin{equation}
\frac{dE}{dt} =\upsilon_{s}\int_{-\infty}^{+\infty}d\xi\left(F\left(m_{R}^{0},\psi_{s}\right)\frac{\partial\psi_{s}^{*}}{\partial\xi}+F^{*}\left(m_{R}^{0},\psi_{s}\right)\frac{\partial\psi_{s}}{\partial\xi}\right)\label{HamiltonianApproach},
\end{equation}
with $F\left(m_{R}\text{,}\psi\right)=\left(\bar{g}_{R}+\frac{i}{2}\bar{R}\right)m_{R}\psi$ and the energy of the dark soliton being given by
\begin{eqnarray}
E&=&\frac{1}{2}\int_{-\infty}^{+\infty}d\xi\left[\left|\frac{\partial\psi_{s}}{\partial\xi}\right|^{2}+\left(1-\left|\psi_{s}\right|^{2}\right)^{2}\right]\nonumber\\
&=&\frac{4}{3}\left(1-\upsilon_{s}^{2}\left(t\right)\right)^{\frac{3}{2}}.\label{SolitonEnergy}
\end{eqnarray}
With the knowledge of the dark soliton's energy of Eq. (\ref{SolitonEnergy}), we can easily calculate the time variation of the soliton's energy, i.e. the left hand of Eq. (\ref{HamiltonianApproach}).
However, the calculations of the right-hand side of Eq. (\ref{HamiltonianApproach}) require the information of the reservoir density of $m_R$.

In this work, we limit ourselves to the experiment-related fast reservoir limit, i.e. in the parameter regime of $\bar{\gamma}_C\ll \bar{\gamma}_R$. In such, directly following Refs. \cite{Smirnov2014,Xu2019}, we can obtain
the density of reservoir as follows
\begin{equation}
m_{R}^{0}\left(\xi\right)=\frac{\bar{\gamma}_{C}\left(1-\left|\psi\right|^{2}\right)+\bar{P}_{\mathcal{P}\mathcal{T}}\left(\xi\right)}{\bar{\gamma}_{R}},\label{FFR}
\end{equation}
with $\bar{P}_{\mathcal{P}\mathcal{T}}$ being the dimensionless form $\mathcal{P}\mathcal{T}$- symmetric pumping of Eq. (\ref{PTpotential}). Substituting Eq. (\ref{FFR}) into Eq. (\ref{Fpsi}), we can obtain
\begin{eqnarray}
i\frac{\partial\psi}{\partial t}	&=&	-\frac{1}{2}\nabla^{2}\psi+\left(1-\frac{\bar{g}_{R}\bar{\gamma}_{C}}{\bar{\gamma}_{R}}\right)\left(\left|\psi\right|^{2}-1\right)\psi\nonumber\\
&+&\frac{\bar{g}_{R}}{\bar{\gamma}_{R}}\bar{P}_{\mathcal{P}\mathcal{T}}\psi+\frac{i}{2}\frac{\bar{R}\bar{\gamma}_{C}}{\bar{\gamma}_{R}}(1-\left|\psi\right|^{2})\psi
+\frac{i}{2}\frac{\bar{P}_{\mathcal{P}\mathcal{T}}}{\bar{\gamma}_{R}}\psi.\label{FinalPT}
\end{eqnarray}
Equation (\ref{FinalPT}) is the key result of this Letter. In detail,
the first term in the second line of Eq. (\ref{FinalPT}) labels the $\mathcal{P}\mathcal{T}$- symmetric potential, giving the $\mathcal{P}\mathcal{T}$- symmetry-induced non-Hermitian; the second term  in the second line of Eq. (\ref{FinalPT}) represents the non-Hermitian due to the intrinsic non-equilibrium nature; the last term in the second line of Eq. (\ref{FinalPT}) can only exist in the case of both $\bar{P}_{\mathcal{P}\mathcal{T}}$ and  $\bar{\gamma}_R$ being non-zero. In such, Equation (\ref{FinalPT}) can capture the combined effects of non-Hermitian due to both nonequilibrium nature and $\mathcal{P}\mathcal{T}$ symmetry in determining the properties of a system by including the nonequilibrium nature characterized the intrinsic kinds of gain and loss of a system, to a $\mathcal{P}\mathcal{T}$- symmetric many-body quantum system.

\begin{figure}
  \includegraphics[width=1.0\columnwidth]{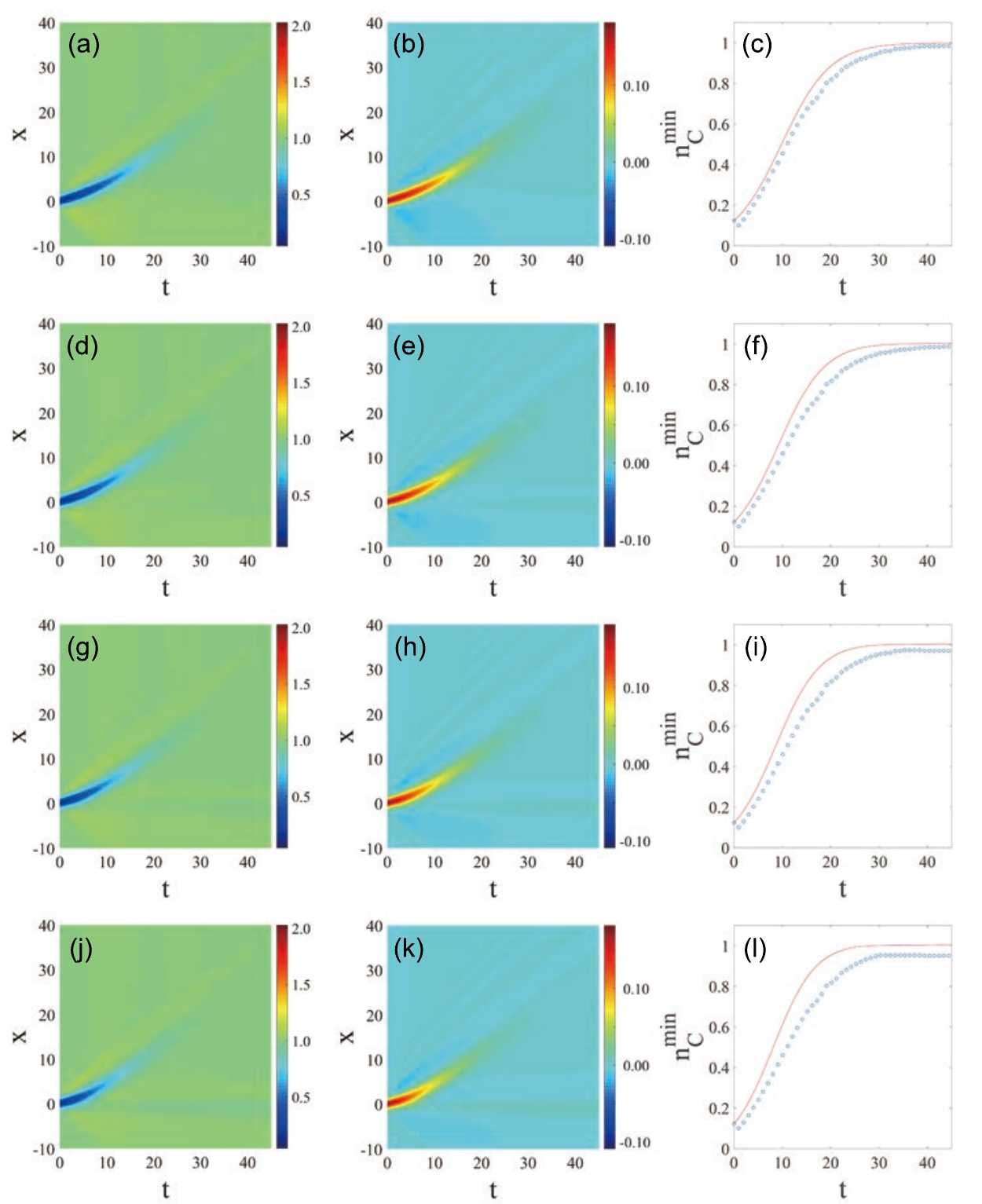}\\
\caption{Dynamics of a 1D dark soliton with the initial velocity $\upsilon_{s0}=0.35$
in the case of weak pumping. Shown are contour plots of $n_{C}$$\left(x,t\right)$
and the real part of $m_{R}\left(x,t\right)$and the dependence $n_{C}^{min}(t)$
computed using Eqs. (15) and (16). Parameters are $\bar{g}_{R}=2$, $\bar{\gamma}_{C}=3$, and $\bar{\gamma}_{R}=15$, $\bar{R}=1.5$, and (a)-(c) $P_{1}=0$, $P_{2}=0$; (d)-(f) $P_{1}=0.01$, $P_{2}=0.3$;
(g)-(i) $P_{1}=0.01$, $P_{2}=0.6$; (j)-(l) $P_{1}=0.01$, $P_{2}=0.9$.
}\label{Fig2}
\end{figure}

Finally, by plugging Eqs. (\ref{SolitonEnergy}) and (\ref{FFR}) into Eq. (\ref{HamiltonianApproach}), one can readily arrive at the equation of motion for the
dark soliton's velocity as follows
\begin{equation}
\frac{d\upsilon_{s}}{dt}  =F_{\text{eff}},\label{VE}
\end{equation}
with $F_{\text{eff}}$ representing an effective force given by
\begin{equation}
F_{\text{eff}}=\frac{1}{3}\frac{\bar{R}\bar{\gamma}_{C}}{\bar{\gamma}_{R}}\upsilon_{s}\left(1-\upsilon_{s}^{2}\right)
+\frac{1}{3}\frac{\bar{R}P_{1}}{\bar{\gamma}_{R}}\upsilon_{s}+\frac{\pi}{32}\frac{\bar{R}P_{2}}{\bar{\gamma}_{R}}\sqrt{1-\upsilon_{s}^{2}}.\label{Force}
\end{equation}
Based on Eq. (\ref{VE}), we can regard the dynamics of the dark soliton as the motion of a classical particle of mass $m_{\text{eff}}=1$ subjected to an external force $F_{\text{eff}}$.
As a first check of validity of Eq. (\ref{VE}), we consider the case of vanishing $\mathcal{P}\mathcal{T}$- symmetric pumping, i.e. $P_1=P_2=0$. In such a case, Equation (\ref{VE}) can exactly cover
the corresponding results in Ref. \cite{Smirnov2014}.

Based on Eq. (\ref{VE}), the effects of $\mathcal{P}\mathcal{T}$-symmetric pumping in the non-equilibrium scenario on the dark soliton can be explained as follows: (i) The first term in Eq. (\ref{Force})
comes from the effects due to the non-equilibrium nature; the second and third terms originate from the real and imaginary part of  $\mathcal{P}\mathcal{T}$-symmetric pumping. (ii) The last two terms in
Eq. (\ref{Force}) contains two competitive parts: while the action of the first part leads to acceleration of the dark soliton, the second part can be rewritten as $\frac{\pi}{32}\frac{\bar{R}P_{2}}{\bar{\gamma}_{R}}(1-\frac{1}{2}\upsilon_{s}^{2})$ in the limit of slow velocity. Then the imaginary part of $\mathcal{P}\mathcal{T}$- symmetric pumping affects the dark soliton by giving a background fixed force of $\frac{\pi}{32}\frac{\bar{R}P_{2}}{\bar{\gamma}_{R}}$ and
 slowing down the motion of the dark soliton with the force of $-\frac{\pi}{64}\frac{\bar{R}P_{2}}{\bar{\gamma}_{R}}\upsilon_{s}^{2}$. With the knowledge of $v_{\text{s}}$ by solving Eq. (\ref{VE}),  we can proceed to obtain determines darkness of a dark soliton through the simple relation $n^{\text{min}}_C\left(t\right)=v^2_{\text{s}}(t)$.

Above, we have developed the analytical physical picture of the dark soliton in a polariton condensate under the nonresonant $\mathcal{P}\mathcal{T}$- symmetric pumping. Below we are ready investigate how the combined effects of
the non-equilibrium nature and $\mathcal{P}\mathcal{T}$- symmetry on a dark soliton (see Figs. \ref{Fig2}) by numerically solving Eqs. (\ref{Fpsi}) and (\ref{FmR}) with the initial condition of Eq. (\ref{ansatz}).

As a benchmark for later analysis, we recall the case of the vanishing $\mathcal{P}\mathcal{T}$- symmetric pumping ($P_1=P_2=0$), corresponding to the scenario of Ref. \cite{Smirnov2014}. As shown in Figs. \ref{Fig2} (a)-(b), the non-equilibrium nature of the model
system is to blend the soliton with the background at a finite time. Meanwhile, trajectory and lifetime predicted by Eq. (\ref{VE}) (see Fig. \ref{Fig2} (c)) are in excellent agreement with direct numerical simulations.
We begin with discussing the effect of $\mathcal{P}\mathcal{T}$- symmetric pumping characterized by the parameters of $P_1$ and $P_2$ on the dark soliton of a polariton condensate with the fixed nonequilibrium nature related parameters.
For this purpose, we devise the following scenario: we fix the real part $P_1$ of $\mathcal{P}\mathcal{T}$- symmetric pumping and investigate the role of the imaginary part $P_2$ on the dark soliton. In Figs. \ref{Fig2}, we show the evolution of the condensate density
distribution $n_{C}\left(x,t\right)=\left|\psi\left(x,t\right)\right|^{2}$and associated perturbations of the polariton reservoir density $m_{R}(x,t)$
in the following figure (left and middle columns, respectively). The right column in figure shows the time dependence of the minimum value
of the condensate density associated with the dark soliton. The red line shows the darkness $n_{C}^{min}(t)=\upsilon_{s}^{2}(t)$ calculated analytically using Eq. (\ref{VE}), and shows an excellent agreement with numerics.
As shown in the left column of Fig. \ref{Fig2}, although the lifetime of the dark soltion becomes to be shorter with the increase of the $P_2$, the corresponding propagation time for  the dark soliton reaches $t\sim 10^2$ ps, which is much longer than the condensate and reservoir relaxation times. Note that with the further increase of $P_2$, it's supposed that there will exist a quantum phase transition from the $\mathcal{P}\mathcal{T}$- symmetry phase to $\mathcal{P}\mathcal{T}$- symmetry broken phase~\cite{Musslimani2008,Lumer2011}.  The dark soliton should be destroyed in the parameter regime of $\mathcal{P}\mathcal{T}$- symmetry broken phase, which is beyond the scope of this work.

In summary, we have investigated the dynamics of dark solitons appearing in a polariton condensates under non-resonant  $\mathcal{P}\mathcal{T}$- symmetric pumping. In particularly, we have derived analytical expression of the equation of motion of the velocity of the dark soliton. Within the framework of Hamiltonian approach, our analytical results capture the essential physics as to how the combined effects of the open-dissipative and $\mathcal{P}\mathcal{T}$ Symmetry affects the lifetime of the dark solitons. We also solve the modified open-dissipative GP equation numerically. The numerical results find agreement with the analytically ones. We also demonstrate that the dark solitons can exist in a long time which being influenced by the $\mathcal{P}\mathcal{T}$- symmetric parameters.

{\it Acknowledgement---} We acknowledge constructive suggestions from Augusto Smerzi, and thank Xingran Xu, Biao Wu, Chao Gao, and Yan Xue for stimulating
discussions. This work is financially supported by the key projects of the Natural Science Foundation of China (Grant No. 11835011).

\bibliography{myr}

\end{document}